\documentclass[a4paper,12pt]{article}
\pdfoutput=1
\usepackage{times}
\usepackage[T1]{fontenc}
\usepackage[cp1250]{inputenc}
\usepackage{amsfonts}
\usepackage{amsmath}
\usepackage{graphics}
\usepackage{color}
\usepackage{epsfig,epsf}

\newcommand{\be}{\begin{equation}}
\newcommand{\ee}{\end{equation}}
\newcommand\beq{\begin{eqnarray}}
\newcommand\eeq{\end{eqnarray}}
\def\as{\alpha_{\mathrm{s}}}
\def\abar{\bar\alpha_{\mathrm{s}}}
\def\Qs{Q_{\mathrm{s}}}
\def\muf{\mu_{\mathrm{F}}}
\def\mufo{\mu_{\mathrm{F,0}}}
\def\FL{F_{\mathrm{L}}}
\def\FT{F_{\mathrm{T}}}
\def\aem{\alpha_{\mathrm{em}}}
\def\CF{C_{\mathrm{F}}}
\def\Nc{N_{\mathrm{c}}}
\def\Ren{\mathrm{Re}}
\def\Imn{\mathrm{Im}}

\begin{document}

\titlepage

\begin{center}
\vspace*{2cm}
{\Large \bf  	
Twist decomposition of non-linear effects in Balitsky--Kovchegov \\[1ex] evolution of proton structure functions} \\
\vspace*{7mm}
Leszek Motyka$^{1}$ and Mariusz Sadzikowski$^{2}$
\\
\vspace*{7mm}
{\it
Jagiellonian University, Institute of Theoretical Physics,\\
\L{}ojasiewicza 11,  30-348 Krak\'{o}w, Poland}
\\[1em]
{\it $^1$ leszek.motyka@uj.edu.pl \\
$^2$ mariusz.sadzikowski@uj.edu.pl \\

\vspace*{5mm}
}

{October 31, 2023}

\end{center}

\vspace*{2ex}

\begin{abstract}
Effects of non-linear small-$x$ evolution of the gluon distribution given by the Balitsky--Kovchegov equation are analyzed within the collinear approximation framework. We perform a twist decomposition of the proton structure functions $F_2$ and $\FL$  obtained from the Balitsky--Kovchegov equation using the Mellin representation of the scattering cross-sections at high energies. In both the structure functions we find strong corrections coming from the non-linear effects in the gluon evolution at twist~2, and strongly suppressed higher twist effects. This implies that unitarization effects of high energy scattering amplitudes are mostly the leading twist effect.
Furthermore we consider the double logarithmic limit of the Balitsky--Kovchegov equation for the collinear gluon distribution, and compare the result to the Gribov--Levin--Ryskin equation. We find that these two equations differ by two powers of the hard scale logarithm for the large scales.
\end{abstract}
\newpage

\section{Introduction and conclusions}

The Electron--Ion Collider will probe deep inelastic scattering (DIS) of electrons on large nuclei at high collision energies \cite{Accardi:2012qut,AbdulKhalek:2021gbh} and this will allow to probe partons that carry a small fraction $x$ of the nucleus momentum. The parton distribution functions grow steeply with decreasing values of $x$, and this growth is driven mostly by gluons. Specifically, the gluon distribution function $g(x,Q^2)$ at the scale $Q^2$ enters the DIS cross sections as $xg(x,Q^2)$, that grows at small $x$ as $x^{-\lambda}$ with $\lambda > 0$. For sufficiently small values of $x$ and sufficiently low scales $Q^2$ this growth has to be slowed down and finally tamed by unitarity corrections, see e.g.\ \cite{Bartels:1992ym,Bartels:1993ih,Kovchegov:1999yj,Kovchegov:1999ua,Bartels:1999aw}. These corrections enter into QCD evolution equations as non-linear terms, that may be interpreted as effects of parton recombination at high parton density regime. At very high densities the gluon production by parton splittings and the gluon recombination are expected to balance, leading a phenomenon called gluon saturation.

 Non-linear corrections to linear evolution equation were studied within the two main QCD frameworks. Within the collinear Dokshitzer--Gribov--Lipatov--Altarelli--Parisi (DGLAP) approach a non-linear evolution equation was proposed by Gribov, Levin and Ryskin (GLR) \cite{Gribov:1983ivg}, and by Mueller and Qiu \cite{Mueller:1985wy}. The unitarity effects, however, are more pronounced when viewed from the perspective of small~$x$ evolution as given by the Balitsky--Fadin--Kuraev--Lipatov (BFKL) framework \cite{Kuraev:1977fs,Balitsky:1978ic,Lipatov:1996ts,Fadin:1998py}. It is so because the BFKL framework is more sensitive to lower momentum scales, while in the DGLAP framework the unitarity effects can be greatly reduced by choosing sufficiently large factorization scale. The BFKL equation is linear. The non-linearity enters to this equation in the form of triple BFKL ladder interaction, called the triple pomeron vertex
\cite{Bartels:1992ym,Bartels:1993ih}. The evolution equation equivalent to BFKL with the triple pomeron interaction was derived by Balitsky \cite{Balitsky:1995ub}  and Kovchegov \cite{Kovchegov:1999yj,Kovchegov:1999ua} --- the Balitsky--Kovchegov (BK) equation. In results of further studies of gluon color correlations in the high density regime at small~$x$,  the framework of Color Glass Condensate (CGC) was developed \cite{McLerran:1993ni,McLerran:1993ka,Gelis:2010nm}, and JIMWLK equation was derived for the target wave function \cite{Jalilian-Marian:1997ubg,Iancu:2000hn,Ferreiro:2001qy,Iancu:2001ad,Kovner:2013ona}.

Theoretical understanding of non-linear evolution equations in QCD is pretty good by now, and a large number of phenomenological studies were perfomed to constrain these effects. Probably the most successful analysis showing strong arguments for gluon saturation was performed by Golec--Biernat and W\"{u}sthoff (GBW) \cite{Golec-Biernat:1998zce,Golec-Biernat:1999qor}. It describes HERA data from the inclusive DIS down to photoproduction limit, and the diffractive DIS data in a simple unified framework, called the GBW saturation model. This model allowed for a simple and efficient extensions to include scale evolution effects \cite{Bartels:2002cj} and to describe also other observables, in particular for exclusive diffractive processes \cite{Kowalski:2003hm,Kowalski:2006hc}. In a more formal approach predictions of non-linear evolution equations were multiply tested against the data, mostly by fitting the solutions of BK equations to data on proton structure functions, see e.g.\ Refs.\ \cite{Kutak:2003bd,Kutak:2004ym,Albacete:2009fh,Iancu:2015joa,Beuf:2020dxl}. Thus, the small~$x$ non-linear evolution equations  are quite successful in phenomenological applications.  Besides this, however, theoretical questions arise on their relation to the powerful collinear framework. In particular, it is interesting to understand how the multiple scattering and gluon recombination effects in the small~$x$ formulation appear in the Operator Product Expansion (OPE) framework --- how they map onto the twist expansion of hadron structure functions, and how they modify the DGLAP evolution equation. The aim of the present paper is to address and study these questions.

Multiple parton exchange in QCD induces higher twist terms in the OPE.  The canonical scaling of twist~$\tau$ contribution to hadron structure function is $(\Lambda / Q)^{\tau-2}$, where $\Lambda$ is a low hadronic scale, and $Q$ is the large DIS scale. Hence the higher twist terms are power suppressed and for sufficiently large $Q^2$ they may be safely neglected. At small~$x$, however, the evolution of higher twist terms is more rapid than of the leading twist term. For instance, for the dominant small~$x$ gluon exchange, the twist~4 contribution to the structure functions at small~$x$ may be estimated as $\sim [xg(x,Q^2)]^2 \Lambda^2 / Q^2$ see e.g.\ \cite{Gribov:1983ivg,Mueller:1985wy,Bukhvostov:1985rn,Bartels:1993it,Bartels:1999xt} to be compared with $\sim xg(x,Q^2)$ twist~2 behavior, where $g(x,Q^2)$ is the collinear gluon distribution. It follows that the strong growth of $xg(x,Q^2)$ at small~$x$ may partially compensate the $1/Q^2$ suppression of the twist~4 term, and the higher twist corrections may become significant, which would affect the quality of twist~2 DGLAP fits of the structure functions. In fact, DGLAP fits to proton structure functions measured at HERA down to $x \sim 2 \cdot 10^{-5}$ for $Q^2 > 1$~GeV$^2$ deteriorate for $Q^2 < 5$~GeV$^2$ both for diffractive and inclusive structure functions \cite{H1:2015ubc,Motyka:2012ty,Harland-Lang:2016yfn,Abt:2016vjh,Motyka:2017xgk}, and inclusion of higher twist corrections at small~$x$ was shown to improve the quality of data description \cite{Motyka:2012ty,Harland-Lang:2016yfn,Abt:2016vjh, Motyka:2017xgk}.

Theoretical analysis of higher twist corrections to the structure functions is not easy. First, the number of relevant operators grows quickly with increasing twist. So does the complexity of evolution equations. Furthermore, in the standard DGLAP approach the initial conditions are fitted to data, and with more operators at higher twist, more information would be needed to constrain the initial conditions for their evolution. Certainly, the currently available data do not provide such information. Hence, it is justified to use simplified models of higher twist contributions, that incorporate crucial features coming from QCD. A very useful guidelines for such models comes from saturation models or from small~$x$ non-linear evolution equations. In the present analysis we use the approach developed in Refs.\ \cite{Bartels:2000hv,Bartels:2009tu,Motyka:2012ty}, at first for a twist decomposition of the proton structure functions in the saturation model, and later extended to the twist decomposition of the BFKL cross sections \cite{Motyka:2014jpa}. The method relies on Mellin transforms of the scattering cross sections and relating singularities in the complex Mellin moment plane to contributions with definite twists. Here we apply this approach to the Balitsky--Kovchegov equation and we perform the twist decomposition of the non-linear correction.

In the analysis we use a solution of the BK equation in the form of a series in non-linearity, proposed by Kovchegov \cite{Kovchegov:1999ua}. We aim to estimate the corrections from non-linearity to the linear approximation in the region where they are moderate, so we truncate the expansion to the first term, that is from a single contribution of non-linearity. This is sufficient to obtain the key conclusions: (1) large corrections to proton structure functions from the non-linearity, that enter at twist~2, and (2) small higher twist corrections induced by the non-linear term. {{These results lead to, perhaps surprising, overall conclusion that non-linear BK corrections in the proton structure functions are basically the leading twist effect. To be specific, we find that the twist~2 correction from BK non-linearity to the BFKL result reaches $-50\%$ in proton structure functions for $Q^2 = 5$~GeV$^2$  already at $x=10^{-3}$. On the other hand, the higher twist corrections from both BFKL and BK are found to be at 1\% level for $F_2$ and below 10\% in $\FL$.  It should be kept in mind, however, that the multiple scattering effects resumed by the BK equation may be probed in a different way in other processes, leading to a possibly different picture of saturation effects.}}

On the top of this, by taking the double logarithmic limit of the BK equation we obtain a non-linear evolution equation for the collinear gluon distribution $xg(x,Q^2)$. This equation resembles the GLR equation, but we find that it takes a different form of the non-linear part. For a meaningful comparison it is necessary to consider the double logarithmic limit. The two classes od logarithms that are resummed correspond to powers of $t = \log(Q^2)$ and powers of $y = \log(x_0/x)$. For the hierarchy $t \gg \abar y \gg 1$ we  find that the non-linear term we obtain from the BK equation is weaker by two powers of $\log(Q^2)$ than the corresponding contribution in the GLR equation. This feature may be traced back to vanishing of the triple BFKL ladder vertex when the collinear ordering is imposed, as found in Ref.\ \cite{Bartels:2007dm}.  This is in full consistency with the collinear evolution equation for quasi-partonic higher twist operators \cite{Bukhvostov:1985rn}, in which the gluonic ladder merging vertex is absent.  Hence, we conclude that at high $Q^2$ effects of the BK  non-linearity are smaller than expected, both in the gluon evolution and in the higher twist contributions. In the region where  $t \sim \abar y \gg 1$, however, the powers of logarithms agree in the GLR equation and in the double logarithmic limit of the BK equation.

Furthermore, we investigate the origin of the strong effects of non-linearity at twist~2. We get the most clear answer by performing an analysis of the unitegrated gluon distribution $f(x,k^2) \simeq k^2 (\partial /\partial k^2) xg(x, k^2)$, that depends on the gluon transverse momentum squared $k^2$. In $f(x,k^2)$ emerging from the BK equation unitarity effects are very strong for $k^2 < \Qs^2(x)$, where the characteristic $x$-dependent momentum scale, that is naturally interpreted as the saturation scale. In the region $k^2 < \Qs^2(x)$, the distribution $f(x,k^2)$ is strongly suppressed. Hence, $\Qs^2(x)$ plays the r\^{o}le of a lower cut-off in the integral $xg(x,Q^2) \simeq \int ^{Q^2} \frac{dk^2}{k^2} f(x,k^2)$, connecting the collinear gluon distribution with its unintegrated form. This is a sizable correction that enters at twist~2. This implies, that the non-linear effects are concentrated at low momentum scales and it is possible to factor them out and absorb in the input value of the gluon distribution $xg(x,\muf^2)$, provided that the factorization scale $\muf^2$ is reasonably larger than the saturation scale  $\Qs^2(x)$. Hence, in order to clearly see non-linear BK effects one should probe the region of $\muf^2 < \Qs^2$. This region is not accessible for DIS on the proton, but may be probed in DIS on large nuclei, e.g.\ at the Electron--Ion Collider.

\section{Proton structure functions from Balitsky--Kovchegov equation}

The total inclusive cross-section $\sigma^{\gamma^\ast A}$ of the virtual photon $\gamma^\ast$ scattering on large nucleus $A$, in the high energy limit,
can be described in terms of the structure functions $\FL, F_2$
\be
\label{structure_functions}
F_{\mathrm{T,L}}(x,Q^2) = \frac{Q^2}{4\pi^2\aem}\sigma^{\gamma^\ast A}_{\mathrm{T,L}} ,\;\;\;F_2 (x,Q^2) = \FT (x,Q^2)+\FL (x,Q^2)
\ee
where
\be
\label{photon-proton-csection}
\sigma^{\gamma^\ast A}_{\mathrm{T,L}}(x,Q^2) = \int \frac{d^2\mathbf{r}}{4\pi} \int_0^1 dz |\psi_{\mathrm{T,L}}(z,r,Q^2)|^2 \sigma_{q\bar{q}}(x,\mathbf{r}) .
\ee
The transverse ($T$) and longitudinal ($L$) photon wavefunctions $\psi_{\mathrm{T,L}}(z,r,Q^2)$ describe probability
amplitudes of the fluctuation of the virtual photon into a quark-antiquark dipole of the transverse
size $r=|\mathbf{r}|$ and a fraction $z$ of the longitudianal light-cone momentum carried by the quark \cite{Nikolaev:1990ja}. The
color dipole scatter over a nucleus with the cross section $\sigma_{q\bar{q}}$ given by an imaginary part of the forward
dipole - nucleon scattering amplitude $N(x,\mathbf{r},\mathbf{b})$
\be
\label{dipole_cross_section}
\sigma_{q\bar{q}}(x,\mathbf{r}) = 2\int d^2\mathbf{b}\; N(x,\mathbf{r},\mathbf{b}) \equiv \sigma_0 N(y,\mathbf{r}),
\ee
where $y=\log(x_{\mathrm{in}}/x)$ is a rapidity variable developed with respect to some initial value $x_{\mathrm{in}}$ \cite{Kovchegov:1999yj}. The
cross section parameter $\sigma_0$ is related to the effective nucleus radius  $\sigma_0 = 2\pi R_A^2$ via integration
over the impact parameter $\mathbf{b}$. In the above equations it is already assumed that the main contribution to scattering comes from perturbative dipoles
located far from the edges of the nucleus, namely that a size of the dipole $r$ in the transverse space is much smaller
then the nucleus radius $R_A$. In this way one neglects a non-trivial dependence of the amplitude on the impact parameter $\vec{b}$,
limiting only to a simple cylindrical geometry of the high energy nucleus.

The rapidity evolution of the amplitude $N(y,\mathbf{r})$ is described by the Balitsky--Kovchegov equation \cite{Balitsky:1995ub, Kovchegov:1999yj}, which in momentum space reads
\be
\frac{\partial \phi (y,k^2_\perp)}{\partial y} = \bar{\alpha}_s\int_0^\infty\frac{dq^2_\perp}{q^2_\perp}
\left\{\frac{q^2_\perp\phi (y,q^2_\perp)-k^2_\perp\phi (y,k^2_\perp)}{|q^2_\perp-k^2_\perp|}+\frac{k^2_\perp\phi (y,k^2_\perp)}{\sqrt{4q^4_\perp+k^2_\perp}}\right\}
-\bar{\alpha}_s \phi^2(y,k^2_\perp)
\ee
where $\bar{\alpha}_s=\as \Nc/\pi$ and $\phi (y,k^2_\perp)$ is the Fourier transform of the amplitude
\be
\label{N_fourier}
\phi (y,k^2_\perp) = \int\frac{d^2r}{2\pi} e^{-i\mathbf{k}_\perp\cdot\mathbf{r}}\frac{N(y,\mathbf{r})}{r^2} .
\ee
The first term in the BK equation is given by the BFKL kernel that can be solved exactly using its eigenfunctions \cite{Lipatov:1996ts}
\be
\phi_0 (y,k^2_\perp) = \sum_{n=-\infty}^\infty \int_{c-i\infty}^{c+i\infty}\frac{d\gamma}{2\pi i}\left(\frac{k^2_\perp}{Q_0^2}\right)^{-\gamma}
C_n(\gamma)\exp\{\bar{\alpha}_s\chi (n,\gamma)y+i n\varphi\}
\ee
where $\chi (n,\gamma)$ are eigenvalues of the BK equation and coefficients $C_n(\gamma)$ are determined by the initial condition. Note that we use the convention in which the standard Mellin moment corresponds to $-\gamma$, and the reference scale in the Mellin transform is $Q_0^2$.
The fundamental Mellin strip is located in the interval $c\in (0,1)$. At large rapidity $y$, which is of main interest in this article,
the dominant contribution is given by $n=0$ eigenvalue
\be
\chi (\gamma)\equiv \chi (0,\gamma) = 2\psi (1) - \psi (1-\gamma)-\psi (\gamma)
\ee
and we adopt this approximation throughout the whole paper. { The non-linear BK equation can be solved using iterative procedure
\cite{Kovchegov:1999ua}. In the Mellin space one can write
\beq
\label{series}
\tilde{\phi} (y,\gamma) &=& \frac{1}{Q_0^2}\int_0^\infty dk^2_\perp \left(\frac{k^2_\perp}{Q_0^2}\right)^{\gamma -1} \phi (y,k^2_\perp) \nonumber\\
\tilde{\phi} (y,\gamma) &=& \sum_{i=0}^\infty \tilde{\phi}_i(y,\gamma),\;\;
\eeq
and the functions $\tilde{\phi}_i$ satisfy the infinite set of the coupled equations:
\beq
\label{BK_iteretion}
\frac{\partial \tilde{\phi}_0 (y,\gamma)}{\partial y} &=& \bar{\alpha}_s \chi(\gamma)\tilde{\phi}_0 (y,\gamma), \\
\frac{\partial \tilde{\phi}_1 (y,\gamma)}{\partial y} &=& \bar{\alpha}_s \chi(\gamma)\tilde{\phi}_1 (y,\gamma)
- 2\pi i\,\bar{\alpha}_s\int_{c_1-i\infty}^{c_1+i\infty}\frac{d\gamma_1}{2\pi i}\int_{c_2-i\infty}^{c_2+i\infty}\frac{d\gamma_2}{2\pi i} \delta(\gamma-\gamma_1-\gamma_2)\tilde{\phi}_0 (y,\gamma_1)\tilde{\phi}_0 (y,\gamma_2), \nonumber\\
\frac{\partial \tilde{\phi}_2 (y,\gamma)}{\partial y} &=& \bar{\alpha}_s \chi(\gamma)\tilde{\phi}_2 (y,\gamma) \nonumber\\
&-& 2\pi i\,\bar{\alpha}_s\int_{c_1-i\infty}^{c_1+i\infty}\frac{d\gamma_1}{2\pi i}\int_{c_2-i\infty}^{c_2+i\infty}\frac{d\gamma_2}{2\pi i} \delta(\gamma-\gamma_1-\gamma_2)(\tilde{\phi}_0 (y,\gamma_1)\tilde{\phi}_1 (y,\gamma_2)+\tilde{\phi}_1 (y,\gamma_1)\tilde{\phi}_0 (y,\gamma_2)),\nonumber\\
.....\nonumber
\eeq
and $0<\Ren (c_1,c_2)<1$. From the structure of the equations one can infer that $\tilde{\phi}_i\sim \tilde{\phi}_0^{i+1}$, where the term
$\tilde{\phi}_0\sim \exp(\alpha\chi y)$. Therefore the amplitude $\tilde{\phi}_i$ describes the process with $i+1$ pomeron exchanges included in
the DIS diagram. The series (\ref{series}) is convergent if $\tilde{\phi}_0(y,k^2_\perp)\ll 1$ which locates the quark transverse momenta within the perturbative domain.
The thorough analysis in \cite{Kovchegov:1999ua} shows that the series is convergent for values
of quark momenta above the saturation scale with possible extension using analytical continuation. Such requirement is consistent with the twist decomposition which assumes that the virtuality of the quark--antiquark pair is large with respect to the typical hadronic scale. The procedure is not applicable below the saturation scale.}
However, our main goal in this paper is to estimate the influence of the non-linear corrections from the BK equation on the lowest twists of the structure functions. Therefore, we limit our calculation to the first order correction. It is a simple exercise to solve the second equation of (\ref{BK_iteretion}) with the result
\beq
\label{BK_correction}
\tilde{\phi}_1(y,\gamma) &=& \int\frac{d\gamma_1}{2\pi i}\frac{d\gamma_2}{2\pi i}\,2\pi i \delta(\gamma-\gamma_1-\gamma_2)C_0(\gamma_1)C_0(\gamma_2) \nonumber\\
&&\frac{\exp{(\bar{\alpha}_s y \chi(\gamma))}-\exp{(\bar{\alpha}_s y \chi(\gamma_1)+\bar{\alpha}_s y \chi(\gamma_2))}}{\chi(\gamma_1)+\chi(\gamma_2)-\chi(\gamma)} ,
\eeq
whereas the first equation gives the BFKL solution for $n=0$ eigenvalue. For further analysis we adopt the
exponential form of the initial conditions
\be
N(y=0,\mathbf{r}) = 1-\exp \left(-r^2 Q_0^2\right)
\label{eq:input}
\ee
which gives
\be
C_0(\gamma) = -\frac{2^{2\gamma -1}\Gamma(\gamma)}{\Gamma(1-\gamma)}\Gamma(-\gamma)
\ee
and the BFKL solution takes the form
\be
\tilde{\phi}_0(y,\gamma) = -\frac{2^{2\gamma -1}\Gamma(\gamma)}{\Gamma(1-\gamma)}\Gamma(-\gamma) e^{\bar{\alpha}_s\chi (\gamma) y} ,
\ee
whereas the Mellin strip is limited to $0<\mathrm{Re} (c)<3/4$. Note that we changed convention for the $Q_0$ parameter of the input fuction $N(y=0,\mathbf{r})$ w.r.t.\  Ref.\ \cite{Motyka:2014jpa}: the present $Q_0$ equals $1/2$ of $Q_0$ used in \cite{Motyka:2014jpa}.

\section{Twist decomposition of the DIS cross section}

\begin{figure}[t]
 \begin{center}
 \includegraphics[width=0.65\textwidth]{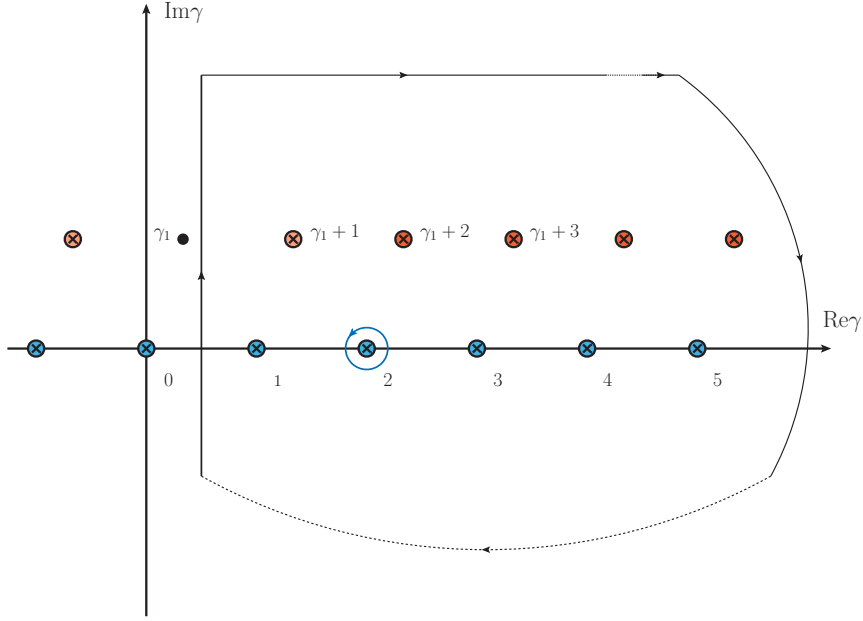}
  \end{center}
 \caption{Location of singularities in a complex Mellin space of $\gamma$ variable. Blue points (on the horizontal axis) correspond to singularities located at integer values, whereas the red ones (above the axis) to integer $\gamma-\gamma_1$. The integration over the large contour is equal
 to the infinite sum of integrals over small contours encircling singular points with a minus sign. One such contour is depicted around point $\gamma=2$.
 }
   \label{Mellin}
\end{figure}

The twist structure of the photon -- nucleus scattering can be obtained from the Mellin transform of the cross section with respect to the virtuality scale. The twist decomposition is performed by isolating contributions of singularities in the complex Mellin moment plane.

The starting point of our analysis is the iterative solution of the BK equation with respect to the non-linear interaction term, as described in the previous section. In this framework the cross section can be described as a series
$$
\sigma^{\gamma^\ast A}_{\mathrm{T,L}} = \sum_{i=0}^\infty \sigma^{(i)\,\gamma^\ast A}_{\mathrm{T,L}},
$$
where
\be
\label{twist_decomp}
\sigma^{(i)\,\gamma^\ast A}_{\mathrm{T,L}}(x,Q^2) = \sigma_0\int_{c-i\infty}^{c+i\infty}\frac{d\gamma}{2\pi i}\left(\frac{4Q_0^2}{Q^2}\right)^{-\gamma}
\tilde{H}_{\mathrm{T,L}} (-\gamma)\frac{\Gamma(1+\gamma)}{2^{-2\gamma-1}\Gamma(-\gamma)}\tilde{\phi}_i (y,-\gamma).
\ee
The Mellin fundamental strip is located in $-3/4<\mathrm{Re}(c) <0$ and $\tilde{\phi}_i$ are solutions of the equations (\ref{BK_iteretion}).
The functions $\tilde{H}_{\mathrm{T,L}}$ are Mellin transforms of the photon wave functions that can be found in \cite{Bartels:2000hv,Bartels:2009tu}. The leading BFKL contribution is given by the formula
\be
\label{sigma_0}
\sigma^{(0)\,\gamma^\ast A}_{\mathrm{T,L}}(x,Q^2) = -\sigma_0\int_{-1/2-i\infty}^{-1/2+i\infty}\frac{d\gamma}{2\pi i} \left(\frac{4Q_0^2}{Q^2}\right)^{-\gamma} \tilde{H}_{\mathrm{T,L}}(-\gamma)
\Gamma (\gamma)e^{\bar{\alpha}_s y \chi(-\gamma)},
\ee
and was described in \cite{Motyka:2014jpa}.
The lowest order correction $\sigma^{(1)\,\gamma^\ast A}_{\mathrm{T,L}}$ to the BFKL cross section follows from
the solution (\ref{BK_correction}) and decomposition (\ref{twist_decomp})
\beq
\label{sigma_1}
\sigma^{(1)\,\gamma^\ast A}_{\mathrm{T,L}} &=& \sigma_0\int_{c-i\infty}^{c+i\infty}\frac{d\gamma}{2\pi i}\left(\frac{4Q_0^2}{Q^2}\right)^{\gamma}
\tilde{H}_{\mathrm{T,L}} (\gamma)\frac{\Gamma(1-\gamma)}{\Gamma(\gamma)}
\int_{c_1-\infty}^{c_1+\infty} \frac{d\gamma_1}{2\pi i}\frac{\Gamma(\gamma_1)\Gamma(\gamma-\gamma_1)}{2\gamma_1(\gamma-\gamma_1)}\times \nonumber\\
&\times& \frac{e^{(\bar{\alpha}_s y \chi(\gamma))}-e^{(\bar{\alpha}_s y \chi(\gamma_1)+\bar{\alpha}_s y \chi(\gamma-\gamma_1))}}{\chi(\gamma_1)+\chi(\gamma-\gamma_1)-\chi(\gamma)}
\eeq
where $0<\mathrm{Re} (c), \mathrm{Re}(c_1)<3/4$. It is important to note that in the
above expression both exponent factors are important to maintain correct analytical structure. { Indeed, the multiples zeros of the denominator in the complex plain are exactly canceled by the contribution from the exponents in the numerator.} The essential singularities are located at integer values of $\gamma, \gamma_1$ and $\gamma-\gamma_1$ (see Fig.\ \ref{Mellin}), therefore the integration over $\gamma$ variable can be decompose into two sums
\beq
\label{sigma_1_sums}
&&\sigma^{(1)\,\gamma^\ast A}_{\mathrm{T,L}} \approx \sum_{n=1}^N \int_{c-i\infty}^{c+i\infty}\frac{d\gamma_1}{2\pi i}\int_{C_n}\frac{d\gamma}{2\pi i}  \tilde I (\gamma,\gamma_1)+
\sum_{n=1}^\infty \int_{c-i\infty}^{c+i\infty}\frac{d\gamma_1}{2\pi i}\int_{C_{n+\gamma_1}}\frac{d\gamma}{2\pi i} \tilde I (\gamma,\gamma_1), \\ \nonumber
&& { I (\gamma,\gamma_1) = \sigma_0\left(\frac{4Q_0^2}{Q^2}\right)^{\gamma}
\tilde{H}_{\mathrm{T,L}} (\gamma)\frac{\Gamma(1-\gamma)}{\Gamma(\gamma)}
\frac{\Gamma(\gamma_1)\Gamma(\gamma-\gamma_1)}{2\gamma_1(\gamma-\gamma_1)}
\frac{e^{(\bar{\alpha}_s y \chi(\gamma))}-e^{(\bar{\alpha}_s y \chi(\gamma_1)+\bar{\alpha}_s y \chi(\gamma-\gamma_1))}}{\chi(\gamma_1)+\chi(\gamma-\gamma_1)-\chi(\gamma)} ,}
\eeq
where $C_w$ is a small clockwise contour located around point $w$ in the complex $\gamma$ plane.
{ The Cauchy theorem that bridges equations (\ref{sigma_1}, \ref{sigma_1_sums}) is satisfied due to the exponential suppression of the large imaginary values $\Imn\gamma$ brought by the photon wavefunctions $H_{\mathrm{T,L}}$.  The asymptotics of $\chi (\gamma)$ does not spoil that in any way, as it consists of two pieces: the $\,-\log(\gamma)$ term that improves the convergence and the $-\pi \cot(\pi \gamma)$ that contributes to the (only) poles of $\chi(\gamma)$ at the integer values of $\gamma$, but is subleading in the other regions. The series (\ref{sigma_1_sums}) is asymptotic \cite{Bartels:2000hv,Bartels:2009tu} and the optimal number $N$ depends
on the values of $(x, Q)$. 
In particular, one can check numerically that for both the transverse and longitudinal cross sections the first two poles provide the agreement with the full result (\ref{sigma_1}) at the level better than one per mile at $Q^2=5$ GeV$^2$ and $x=10^{-3}$. Additionally, inclusion of the second pole on the top of the first improve the result by more than the order of magnitude, which shows the convergence of the expansion for the first two twists.}

The integral $d\gamma$ over $C_n$ from the first term in (\ref{sigma_1_sums}) gives
a direct contribution to twist $\tau=2n$ after integration over $d\gamma_1$ along the line parallel to imaginary
axis within the Mellin strip. A similar integration over $C_{n+\gamma_1}$ from the second term gives contributions to all twists
of order $\tau \geq 2n$ starting with the lowest value $\tau=4$. This fact follows from the integration over $d\gamma_1$ variable.
However, the twists higher then $\tau=4$ are strongly suppressed, therefore one can assign the contribution of
the second integral to twist $\tau=4$ only, with a small error of order 1 per cent or less.
{ Summarising, in numerical calculations, the BK correction to the leading twist was calculated using the first term of (\ref{sigma_1_sums})
with $n=1$ only, whereas the correction to twist $\tau=4$ by the first term with $n=2$ and the second term with $n=1$.}

\section{Results}

\subsection{Numerical results}

\begin{figure}[t]
\begin{center}
\includegraphics[width=0.46\textwidth]{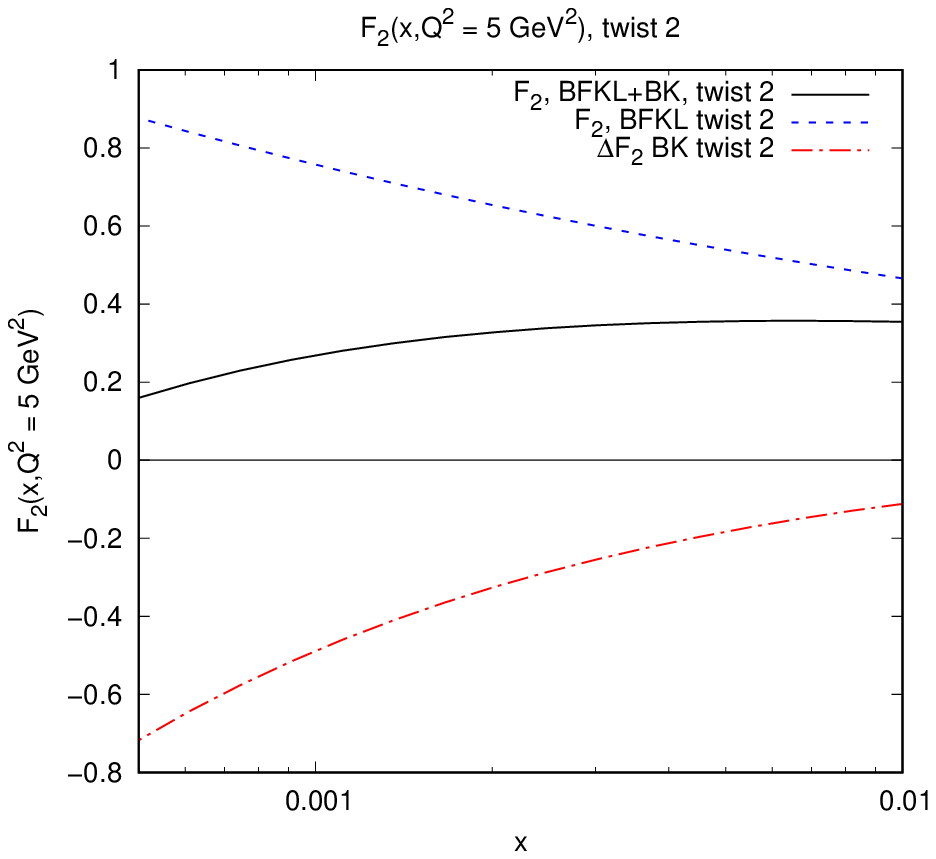} \hspace{5mm}
\includegraphics[width=0.46\textwidth]{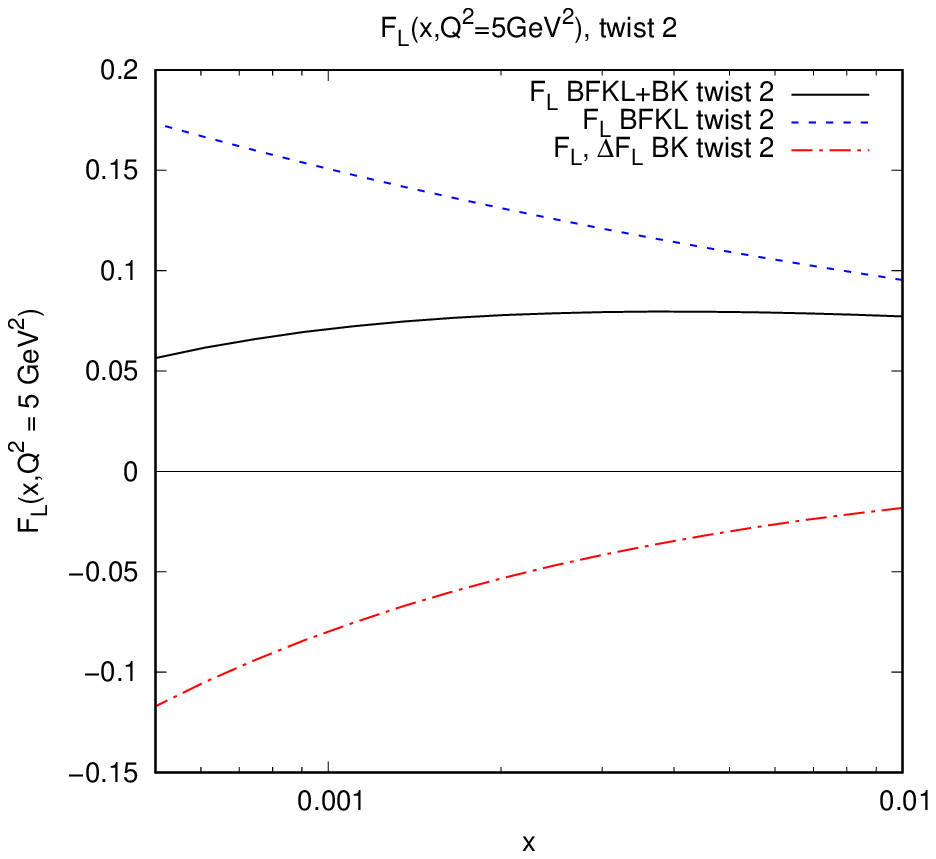} \\
\end{center}
\caption{Effects of non-linear corrections in proton structure functions $F_2(x,Q^2)$ (left) and $\FL(x,Q^2)$ (right) at $Q^2 = 5~\mbox{GeV}^2$ in twist 2 approximation. We show the twist 2 components obtained from the BFKL evolution, the BK correction and their sum.  \label{fig:twist2}}
\end{figure}

With the framework described above we calculate higher twist corrections to the unpolorized proton structure functions $F_2$ and $\FL$ at small $x$. The primary goal is to estimate deviations from linear evolution (BFKL) regime due to non-linearity introduced by the triple Pomeron interaction. We compute the first correction in non-linearity in the iterative solution of Balitsky--Kovchegov equation  --- which we shall call the BK correction. Moreover, we perform an explicit twist decomposition of the structure functions for BFKL result with the BK correction and compare it to the known results computed earlier in the BFKL approach.

As the reference we take the structure functions obtained from a solution of the  leading logarithmic   BFKL equation with parameters obtained in Ref.\ \cite{Motyka:2014jpa}. Let us remind that the input for the BFKL evolution in the dipole representation at $x_{\mathrm{in}} = 0.1$ is assumed to take the GBW form: $\sigma(x_{\mathrm{in}},r) = \sigma_0 [1 - \exp(-r^2 Q_0^2)],$ with $\sigma_0 =17.04$~mb and $Q_0 = 0.255$~GeV. The value of the strong coupling constant $\abar$ in the BFKL kernel is set to $0.087$. This should be understood as an effective value of $\abar$ that partially absorbs the higher order corrections to the BFKL kernel, known to reduce strongly the BFKL Pomeron intercept. This is consistent with application of the Brodsky--Lepage--Mackenzie scale fixing procedure \cite{Brodsky:1982gc} to the NLL BFKL kernel \cite{Brodsky:1998kn}. We stress, however, that the primary goal of the present study is to understand importance of non-linear corrections to BFKL evolution and its twist decomposition and fine details of the model should not affect the key, general features of the results.

We start the numerical analysis from evaluating the BK correction to the BFKL evolution. We compute the leading twist 2 contributions to structure functions $F_2(x,Q^2)$ and $\FL(x,Q^2)$.
We choose the DIS reference scale $Q^2 = 5~\mbox{GeV}^2$, below which the DGLAP fit deteriorates of the final HERA data on structure functions. As it will be clear from the next part of the analysis, the higher twist corrections are small and do not change the conclusions of this part.
In Fig.\ \ref{fig:twist2} we display the twist~2 contributions to the structure functions from: the BFKL equation,  the BK correction and from the sum of BFKL and BK parts. Both the BFKL part and the BK correction are obtained with the same input. The BK corrections to both structure function are large and negative. The magnitude of the corrections, both absolute and relative, grows with decreasing $x$. Clearly, when the relative correction is not small, the higher order corrections of the non-linearity would be necessary to achieve a good approximation of the complete solution of the Balitsky--Kovchegov equation. For the present study it is sufficient  to conclude that the non-linear corrections to BFKL results at twist~2 are large already at $x=0.001$.

\begin{figure}[t]
\begin{center}
\includegraphics[width=0.46\textwidth]{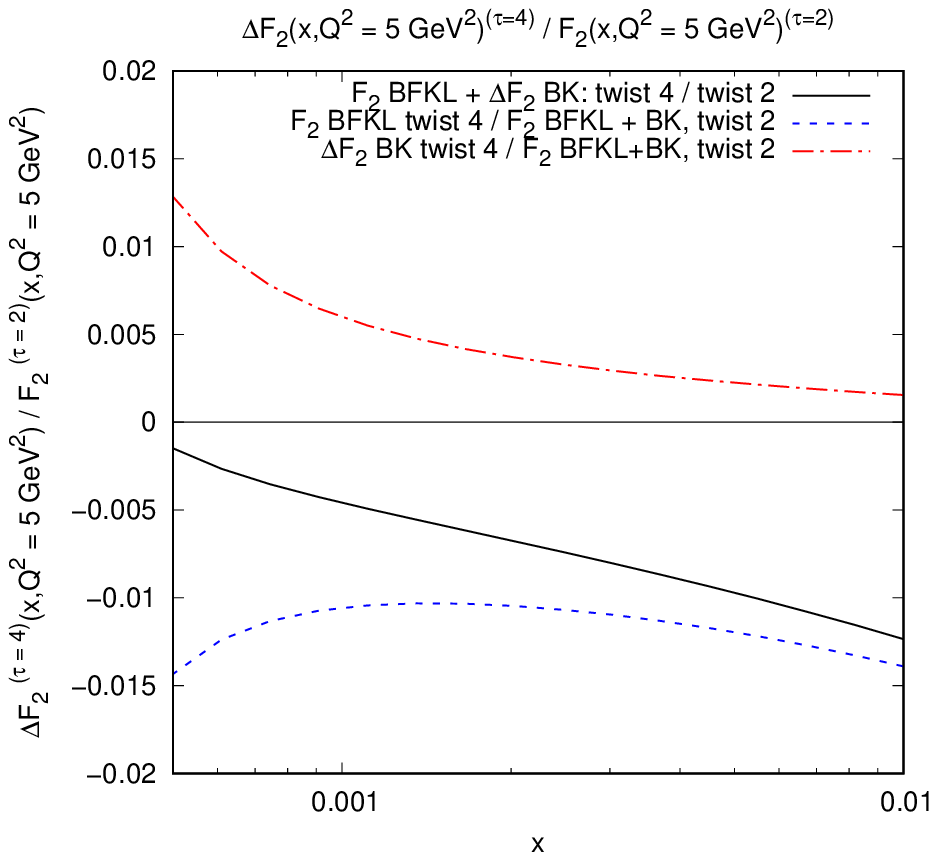} \hspace{5mm}
\includegraphics[width=0.46\textwidth]{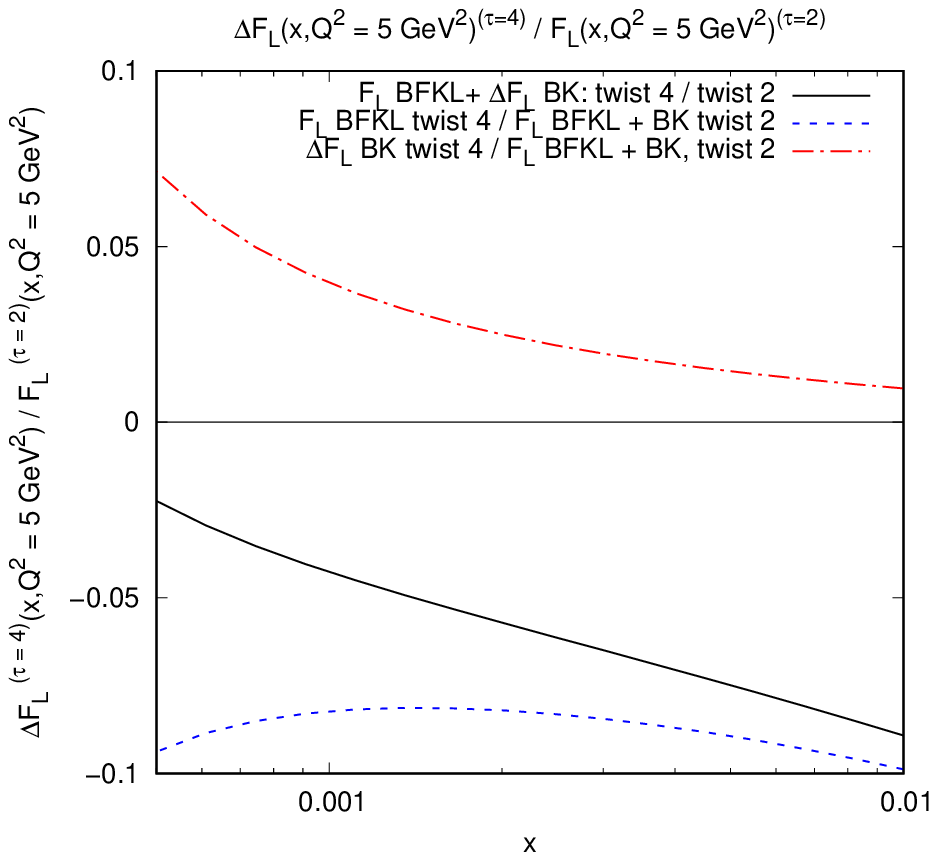} \\
\end{center}
\caption{Relative effects of twist 4 corrections in proton structure functions $F_2(x,Q^2)$ (left) and $\FL(x,Q^2)$ (right) at $Q^2 = 5~\mbox{GeV}^2$. We show the ratios of the twist~4 corrections obtained from the BFKL evolution, the BK correction and their sum to the twist~2 BFKL + BK result.  \label{fig:HT}}
\end{figure}

Next we turn to the analysis of higher twist corrections to the proton structure functions. The most important measure of higher twist effects is its relative magnitude to the twist~2 approximation. We choose as a reference the twist~2 estimates of the structure functions obtained from BFKL equation with the BK correction. We restrict the analysis to twist $\tau=4$ effects. As it is clearly seen in Fig.\ \ref{fig:HT}, the relative twist~4 effects are small or moderate, so it is not necessary to consider the higher, $\tau>4$, corrections.  The magnitudes of relative corrections are different, but the general pattern is similar for both the structure functions. The twist~4 corrections coming from BFKL are negative, and the BK twist~4 contributions are positive, but with smaller absolute value than the leading BFKL contribution. The overall higher twist corrections are negative in both structure functions.
The relative higher twist corrections are found to be larger for $\FL$, where they reach up to about (negative) 10\%. For $F_2$ we find corrections up to about (negative) $1.5\%$. This is expected, as the coefficient function for $\FL$ generates a scale logarithm for twist~4, and does not for twist~2, while for $\FT$, that is dominant in $F_2$, the twist~4 contribution carries one less power of the scale logarithm than the leading twist~2 term \cite{Bartels:2000hv,Bartels:2009tu}. One should keep in mind, however, that the twist content of the cross sections depends not only on the evolution equation, but also on the form of the input, and this dependence is stronger when $x$~is  not very small.

{
The presented results are obtained taking the first order corrections in non-linearity. Most likely this leads to overestimating the non-linear effects. It is expected to happen because the expansion in non-linearity produces the alternating series for the dipole scattering amplitude, as demonstrated in [6]. Effects of the higher orders in non-linearity should turn in when the relative first order correction stops being much smaller than one, and this is certainly the case when the first order correction reaches 50\%. We expect that at higher orders the correction should be somewhat smaller, but this not changes the overall pattern. }

{
We performed a similar analysis also for $Q^2=2$~GeV$^2$ and for $Q^2=10$~GeV$^2$ and found that the general pattern is very similar to the case $Q^2=5$ GeV$^2$. Therefore, we do not depict them in separate figures. Instead, the numerical values of non-linear corrections at twist~2 are given in Table~1 for these three values of $Q^2$ at $x=0.01$ and $x=0.001$. To be specific, we provide
the numerical values of BK corrections $\Delta F_{\mathrm{2,BK}}$,   $\Delta F_{\mathrm{L,BK}}$ compared to the BFKL results in the leading twist for $F_2$ and $\FL$ structure functions.

\begin{table}
  \centering
 \begin{tabular}{|c|c|c|c|}
   \hline
   $x=0.01$ & $Q^2=2$ GeV$^2$ & $Q^2=5$ GeV$^2$ & $Q^2=10$ GeV$^2$ \\
   \hline
    $\Delta F_{\mathrm{2,BK}}/ F_{\mathrm{2,BFKL}}$
     & -26$\%$ & -24$\%$ & -22$\%$ \\
    $\Delta F_{\mathrm{L,BK}}/ F_{\mathrm{L,BFKL}}$
   & -21$\%$ & -19$\%$ & -18$\%$ \\
   \hline
   $x=0.001$ &   &   &  \\
   \hline
    $\Delta F_{\mathrm{2,BK}}/ F_{\mathrm{2,BFKL}}$
   & -73$\%$ & -64$\%$ & -59$\%$ \\
$\Delta F_{\mathrm{L,BK}}/ F_{\mathrm{L,BFKL}}$
   & -60$\%$ & -53$\%$ & -48$\%$ \\
   \hline
 \end{tabular}
  \caption{Contribution of the BK corrections to structure functions $F_2, \FL$ at the leading twist.}
\end{table}

In Table 2 one can find the ratio of twist 4 and twist 2 for $F_2$ ($R_{F_2}$) and $\FL$ ($R_{\FL}$)  structure functions:
$R_{F_2} = (F_{\mathrm{2,BFKL}}^{(4)}+\Delta F_{\mathrm{2,BK}}^{(4)})/(F_{\mathrm{2,BFKL}}^{(2)}+\Delta F_{\mathrm{2,BK}}^{(2)})$,
$R_{\FL} = (F_{\mathrm{L,BFKL}}^{(4)}+\Delta F_{\mathrm{L,BK}}^{(4)})/(F_{\mathrm{L,BFKL}}^{(2)}+\Delta F_{\mathrm{L,BK}}^{(2)})$,
as well as, the ratio of BK correction twist 4 to twist 2 for both structure functions:
$R_{\Delta F_2} = \Delta F_{\mathrm{2,BK}}^{(4)}/(F_{\mathrm{2,BFKL}}^{(2)}+\Delta F_{\mathrm{2,BK}}^{(2)})$, $R_{\Delta \FL} = \Delta F_{\mathrm{L,BK}}^{(4)}/(F_{\mathrm{L,BFKL}}^{(2)}+\Delta F_{\mathrm{L,BK}}^{(2)})$.

\begin{table}
  \centering
 \begin{tabular}{|c|c|c|c|}
   \hline
   $x=0.01$         & $Q^2=2$ GeV$^2$ & $Q^2=5$ GeV$^2$ & $Q^2=10$ GeV$^2$ \\
   \hline
   $R_{F_2}$        & -2.2$\%$ & -1.2$\%$ & -0.7$\%$ \\
   $R_{\Delta F_2}$ & 0.4$\%$ & 0.15$\%$ & 0.08$\%$ \\
   $R_{\FL}$        & -17.4$\%$ & -8.9$\%$ & -5.1$\%$ \\
   $R_{\Delta \FL}$ & 2.1$\%$ & 1.0$\%$ & 0.5$\%$ \\
   \hline
   $x=0.001$        &   &   &  \\
   \hline
   $R_{F_2}$        & -0.2$\%$ & -0.5$\%$ & -0.3$\%$ \\
   $R_{\Delta F_2}$ & 1.6$\%$ & 0.6$\%$ & 0.3$\%$ \\
   $R_{\FL}$        & -8.6$\%$ & -4.3$\%$ & -2.4$\%$ \\
   $R_{\Delta \FL}$ & 10.3$\%$ & 4.0$\%$ & 2.0$\%$ \\
    \hline
 \end{tabular}
  \caption{Ratio of twist 4 to twist 2 for structure functions $F_2, \FL$ and non-linear corrections. Definitions can be found in the main text.}
\end{table}
The overall pattern seen in both the tables agrees with expectations. The effects of non-linearity are strongest at the leading twist, they grow with decreasing~$x$ and slightly decrease with $Q^2$. The higher twist corrections due to non-linearity behave in a similar way, but they are much smaller and their decrease with $Q^2$ is much faster. Let us mention that the results for $Q^2=2$ GeV$^2$ are presented here rather for completeness than for phenomenological purposes. At such low values of $Q^2$ and low $x$ the non-linear corrections are of the same order as the linear part. Therefore, accurate numerical predictions for sure require higher terms of the expansion in non-linearity.}

\subsection{Discussion}

The obtained results show a stronger dependence on $x$ of the twist~4 BK correction than the twist~$4$ BFKL contribution. This is an expected results. In an earlier study \cite{Motyka:2014jpa} we found that in the double logaritmic saddle point approximation the rapidity dependence of the twist~4 BFKL term is governed by $\exp\left(2\sqrt{\abar y \log(Q^2/Q_0^2)}-2\as y\right)$ (up to power factors of $y$), to be compared with twist~2 BFKL in the same approximation $\sim \exp\left( 2\sqrt{\abar y \log(Q^2/Q_0^2}\right)$. This means that in the leading logarithmic BFKL evolution terms corresponding to a double gluon ladder exchange in the total cross section, expected to grow as $\sim \left[\exp\left(2\sqrt{\abar y \log(Q^2/Q_0^2)}\right)\right]^2$, are absent. In other words the multiple elementary $t$-channel gluons present in the reggeized gluons of the BFKL formalism have zero projection on the leading twist~4 exchange in the double logarithmic approximation. The BK correction is different because it is generated by the triple gluon ladder vertex, that couples the genuine two gluon ladder contribution to  the BFKL evolved $\gamma^*$ scattering. Therefore one expects the strong $\sim  \left[\exp\left(2\sqrt{\abar y \log(Q^2/Q_0^2)}\right)\right]^2$ growth of the BK twist~4 correction at asymptotically large $y$ and its dominance in total twist~4 at very small~$x$. The presented results indicate however, that this asymptotic regime is not reached in HERA kinematics.

The overall picture emerging from the numerical analysis is quite clear. It turns out that the non-linear evolution, as given by the BK equation, have strong effects in the leading twist~2 component of the structure functions, and the higher twist components coming from the BK equation are strongly suppressed w.r.t.\ the leading twist terms. It should be stressed that the large non-linear corrections found in BK at twist~2 affect the results obtained in the BFKL framework, and  the corresponding effect in the DGLAP framework depends on the factorization scale, as discussed below. Combining the large BK effects in twist~2 and weak at higher twists, we conclude that the non-linear corrections are concentrated at low scales. Hence we expect that in DGLAP framework the BK corrections can be mostly absorbed into the input for the twist~2 gluon evolution, provided the initial scale of DGLAP evolution $\mufo$ is big enough. Clearly, the scale for the BK effects is the saturation scale $\Qs$, so we conclude that with $\mufo \gg \Qs$  the DGLAP description should not be significantly affected by non-linear evolution effects. This conclusion may change when $\mufo < \Qs(x)$ for some range of $x$ probed by the data.  Then one would expect a significant modification the twist~2 DGLAP evolution due to non-linearity. Fits to the proton structure functions at HERA assuming saturation effects indicate that in HERA kinematics $\Qs^2 < 1$~GeV$^2$. Hence, with DGLAP fits assuming typically $\mufo^2 = 2$~GeV$^2$ or a higher value, they should not be affected by non-linearity. The situation may change, however, for DIS on a large nucleus with the mass number $A$, for which the saturation scale $\Qs^2$ is enhanced by $A^{1/3}$ w.r.t.\ the saturation scale in proton. In order to account for the possibility of $\Qs > Q_0$, in the next section we shall consider the effects of non-linearity on the $Q^2$ evolution in this regime.

\section{Non-linear evolution in the collinear approximation}

\subsection{The double logarithmic regime}
\label{Sec:NL-DLA}

In the numerical analysis we found strong effects of the BK correction in twist~2 components of the structure functions. In order to better understand the origin of this effect let us consider the double logarithmic limit of the BK equation. It is convenient to start from the BK evolution for the unintegrated gluon distribution, $f(x,k^2)$, related to the collinear gluon distribution by the LL formula, $xg(x,Q^2) = \int ^{Q^2} dk^2 f(x,k^2) / k^2$.
In what follows we shall also use notation $f(x,k^2) \to f(y,k^2)$ with $y= \log(x_{\mathrm{in}}/x)$. We keep the convention for the Mellin transform used in the previous sections:
\be
\tilde f(y,\gamma) = \int { dk^2 \over k^2} f(y,k^2) (k^2 / Q_0^2)^{\gamma}, \qquad\qquad
f(y,k^2) = \int_{c - i \infty} ^{c + i\infty} {d\gamma \over 2\pi i} \tilde f(y,\gamma) (k^2 / Q_0^2)^{-\gamma}.
\ee
{ The unintegrated gluon distribution is related to the dipole amplitude through the formula \cite{Bondarenko:2006ft}
\be
f(x,k^2) = \frac{N_c S_{\perp}}{4\alpha_s\pi^2}(k^2)^2\nabla^2_k\phi (y,k^2)|_{y = \log(1/x)} ,
\ee
where the function $\phi (y,k^2)$ is the Fourier transform of $N(y,\mathbf{r})/r^2$, see Eq.\ (\ref{N_fourier}).} The BK equation for the unintegrated gluon distribution reads \cite{Kutak:2003bd,Bondarenko:2006ft}:
\[
{\partial f(y,k^2 ) \over \partial y} = \abar k^2 \int {da^2 \over a^2}
\left[ {f(y,a^2) - f(y,k^2) \over |a^2 - k^2|} +{ f(y,k^2) \over \sqrt{4a^4 + k^4}} \right]
\]
\be
-{2 \pi \as^2 \over S_{\perp}}
\left[k^2\,\int_{k^2} {da^2 \over a^4} f(y,a^2) \int_{k^2} {db^2 \over b^4} f(y,b^2)
\, + \,
f(y,k^2) \int_{k^2} {da^2 \over a^4} \log(a^2 / k^2) \,f(y,a^2)
\right],
\label{eq:BK-momentum}
\ee
where $S_{\perp}$ is the transverse target area, for a uniform target with radius $R_A$, $S_{\perp} = \pi R_A^2$. The first line represents the linear part (the BFKL equation), and the second line is a non-linear correction corresponding to the triple pomeron interaction in the BK equation. Note that the non-linear term corresponds to integrals with a strict anti-collinear ordering, $a^2, b^2 > k^2$, so it vanishes in the collinear limit.  In the Mellin representation this equation reads:
\[
{\partial \tilde f(y,\gamma) \over \partial y} =
\abar \chi (-\gamma)\tilde f(y,\gamma)
\]
\be
-{2 \pi \as^2 \over S_{\perp}Q_0 ^2}
\int {d\gamma_1 \over 2\pi i}\int {d\gamma_2 \over 2\pi i} \, 2\pi i \, \delta(\gamma_1 + \gamma_2 + 1- \gamma ) \, \left[ {1 \over (\gamma_1 + 1) (\gamma_2 +1 ) }+ {1 \over (\gamma_1 + 1)^2} \right] \, \tilde f(y,\gamma_1)  \tilde f(y,\gamma_2),
\ee
which, using the symmetry between $\gamma_1$ and $\gamma_2$ can be rewritten as
\[
{\partial \tilde f(y,\gamma) \over \partial y} =
\abar \chi (-\gamma)\tilde f(y,\gamma)
\]
\be
-{ \pi \as^2 \over S_{\perp} Q_0 ^2}
\int {d\gamma_1 \over 2\pi i}\int {d\gamma_2 \over 2\pi i} \,
2\pi i \, \delta(\gamma_1 + \gamma_2 + 1- \gamma ) \,
{(\gamma+1)^2 \over (\gamma_1 + 1)^2 (\gamma_2 +1 )^2 } \,
\tilde f(y,\gamma_1)  \tilde f(y,\gamma_2).
\ee
This implies the evolution equation for collinear gluon distribution:
\[
{\partial \tilde g(y,\gamma) \over \partial y} =
\abar \chi (-\gamma)\tilde g(y,\gamma)
\]
\be
+{ \pi \as^2 \over S_{\perp} Q_0^2}
\int {d\gamma_1 \over 2\pi i}\int {d\gamma_2 \over 2\pi i} \,
2\pi i \, \delta(\gamma_1 + \gamma_2 + 1- \gamma ) \,
{(\gamma+1)^2 \over (\gamma_1 + 1)^2 (\gamma_2 +1 )^2 } \,
{\gamma_1 \tilde g(y,\gamma_1) \, \gamma_2 \tilde g(y,\gamma_2) \over \gamma},
\ee
where $\tilde g(y,\gamma) = \left. \int d Q^2 / Q^2 \, xg(x,Q^2) (Q^2/Q_0^2)^{\gamma}\right|_{x = x_{\mathrm{in}} \exp(-y)}$ is the collinear gluon distribution $xg(x,Q^2)$ in the Mellin representation. Note that the relation $f(x,k^2) = k^2 \partial_{k^2} xg(x,k^2)$ leads to the $\tilde f(y,\gamma) = (-\gamma)\tilde g(y,\gamma)$ relation in the $(y,\gamma)$ variables.
In the double logaritmic limit, which corresponds to the leading powers of $\gamma$ around $\gamma = 0$, we obtain:
\[
(-\gamma) {\partial \tilde g(y,\gamma) \over \partial y} =
\abar \tilde g(y,\gamma)
\]
\be
-{ \pi \as^2 \over S_{\perp} Q_0^2}
\int {d\gamma_1 \over 2\pi i}\int {d\gamma_2 \over 2\pi i} \,
2\pi i \, \delta(\gamma_1 + \gamma_2 + 1- \gamma ) \,
{(\gamma+1)^2 \over (\gamma_1 + 1)^2 (\gamma_2 +1 )^2 } \,
{\gamma_1 \tilde g(y,\gamma_1) \, \gamma_2 \tilde g(y,\gamma_2)},
\label{eq:BK-glr}
\ee
where we approximated $\chi(-\gamma) \simeq -1/\gamma + {\cal O}(1)$ around $\gamma \to 0$.
{ Before further analysis of this equation let us compare it to the GLR equation~\cite{Gribov:1983ivg,Kovchegov:2012mbw}:
\begin{equation}
\frac{\partial^2 xg(x,Q^2)}{\partial y\partial\log (Q^2/Q_0^2)} = \bar{\alpha}_s xg(x,Q^2)-{\Nc \over 2 \CF}{ \pi \as^2 \over S_{\perp} Q^2}(xg(x,Q^2))^2
\end{equation}
which in the Mellin representation $(y,\gamma)$ takes the form:
\[
(-\gamma) {\partial \tilde g(y,\gamma) \over \partial y} =
\abar \tilde g(y,\gamma)
\]
\be
-{\Nc \over 2 \CF}{ \pi \as^2 \over S_{\perp} Q_0^2}
\int {d\gamma_1 \over 2\pi i}\int {d\gamma_2 \over 2\pi i} \,
2\pi i \, \delta(\gamma_1 + \gamma_2 + 1- \gamma ) \,
\tilde g(y,\gamma_1) \tilde g(y,\gamma_2),
\ee
with $\CF = (\Nc^2 - 1)/2\Nc$. Let us notice that after the Mellin transform the non-linear term becomes the convolution as expected, and the $1/Q^2$ factor changes into the $1/Q_0^2$ which carries the physical dimension.} We find differences between the non-linear equation (\ref{eq:BK-glr}) and the GLR equation. The color prefactor of GLR $\Nc / 2 \CF = \Nc^2 / (\Nc ^2 -1)$ differs from our by ${\cal  O}(1/\Nc^2)$ and this is beyond the leading $\Nc$ accuracy of the BK equation. There is, however,  a more important difference in the integral kernel in the Mellin space. GLR gives
\be
\tilde K _{\mathrm{GLR}} =  2\pi i \, \delta(\gamma_1 + \gamma_2 + 1- \gamma ),
\label{eq:kerGLR}
\ee
while we obtain:
\be
\tilde K _{\mathrm{BK}} = 2\pi i \, \delta(\gamma_1 + \gamma_2 + 1- \gamma ) \,
{\gamma_1  \gamma_2 (\gamma+1)^2 \over (\gamma_1 + 1)^2 (\gamma_2 +1 )^2 }.
\label{eq:kerBK}
\ee
In order to compare the essential properties of these two kernels and resulting evolution equations let us consider their large $Q^2$ behavior. In this region the evolution is dominated by linear term. Further, we want to determine the leading powers of the logarithm $\log(Q^2/Q_0^2)$ that emerge from both the kernels.

At first, let us focus on the limit $t= \log(Q^2 / Q_0^2) \gg \abar y \gg 1$. In this region the saddle point solution of the linear evolution equation is dominated by the anomalous dimension $\gamma_s = \sqrt{\abar y / t} \ll 1$. Note that in the convention for the Mellin moments applied in this paper, they equal to minus anomalous dimensions.

As already said, in the large $Q^2$ regime, the solution for the gluon distribution  $\tilde g(y,\gamma_i)$ is dominated by $\gamma_i \sim 0$, so the Dirac~$\delta$ imposes $\gamma \simeq 1$ in the non-linear correction term (\ref{eq:kerBK}). Therefore it may be approximated by
\be
\tilde K _{\mathrm{BK}} \simeq 2\pi i \, \delta(\gamma_1 + \gamma_2 + 1- \gamma ) \,
{4 \gamma_1  \gamma_2}.
\label{eq:kerBKapprox}
\ee
{
It is important to notice that the factor  $4\gamma_1\gamma_2$ in the integral kernel does not change the dominance of the $\gamma_i \sim 0$ region in the integrand. It is because in this region the collinear gluon distributions are strongly enhanced, they behave as: $\tilde g(y,\gamma_i)  \sim \exp(-\abar y / \gamma_i)$, and the prefactors $\gamma_i$ are subleading with respect to the exponentiated $\gamma_i$~pole part.
}
The aforementioned condition $\gamma \simeq 1$ yields the leading $1/Q^2$ dependence of the non-linear correction term, as in the case of the GLR equation, while the leading part of the gluon distribution is given by the linear evolution and remains localized in the region of $\gamma \simeq 0$.
The integral operator corresponding to this kernel has the same structure as the GLR correction term induced by (\ref{eq:kerGLR}), but with the replacement: $\tilde g(y,\gamma_i) \to \gamma_i \tilde g(y,\gamma_i)$.
{ In the double logarithmic regime we get the following integral representation of the linear rapidity evolution equation:}
\be
-\gamma\tilde g(y,\gamma) \simeq \abar \int^y dy' g(y',\gamma).
\ee
{
This implies that the factors $\gamma_i\tilde g(y,\gamma_i )$ that appear in the non-linear term of Eq.\ (\ref{eq:BK-glr}) are suppressed by one order of $\abar$ in the resummations of scale logarithms.  Another way to see this is to use the saddle point solution to the gluon evolution equation, the solution, \[
xg(x,Q^2) = \left. A(\abar y / \log^3(Q^2)) ^{1/4} \exp(2\sqrt{\abar y \log(Q^2)}) \right|_{y=\log(x_{\mathrm in} /x)}.
\]
 Multiplication by the Mellin moment $\gamma$ corresponds to the differentiation w.r.t.\ $\log(Q^2)$. It leads to lowering the power of logarithm. Explicitly:
\be
{\partial  xg(x,Q^2) \over \partial \log(Q^2)}  =  \left(\sqrt{\abar y \over \log(Q^2) }- {3 \over 4\log(Q^2)} \right)  xg(x,Q^2).
\ee
In the counting of leading logarithms the relative powers of $\as$ and $\log(k^2)$ are important, and both the factors  $\sqrt{\abar y/ \log(Q^2)} $ and $-3/(4\log(Q^2))$ lower the power of logarithm by one w.r.t.\ the power of $\abar$. }
It follows that one iteration of the non-linear term from the BK equation expressed in terms of the collinear gluon distribution, comes at the $\as ^4 \log(Q^2/Q^2_0)/Q^2$ order, to be compared with the GLR non-linear term, $\sim  \as ^2 \log(Q^2/Q^2_0) / Q^2$.

The last conclusion holds true for the collinear gluon distribution resulting from the resummation of term enhanced by powers of $\log(Q^2/Q^2_0)$, and for $t= \log(Q^2/Q^2_0) \gg \abar y$. For the other asymptotic regions: $\abar y \gg  t \gg 1$, and  $\abar y \sim t \gg 1$ , the dominant value of the anomalous dimension $\gamma_s = \sqrt{\abar y / t}$ is not bounded to be much smaller than one, and there is no significant value reduction in the transition from the gluon collinear gluon distribution function $\tilde g(y,\gamma)$ to the unintegrated one, $\tilde f(y,\gamma) \simeq  \gamma \tilde g(y,\gamma)$, as the loss of one power of the scale logarithm is compensated by a similar or larger enhancement by the factor of $\abar y$. In these regions we recover the logaritmic scaling of the GLR equation.

The conclusion about the strong suppression of the non-linear BK term in the double logarithmic approximation with $t \gg \abar y$ hierarchy can be checked by a direct analysis of this term in momentum space. The transverse momentum integrals in Eq.\ (\ref{eq:BK-momentum}) have lower boundary of $k^2$ --- it corresponds to the anticollinear ordering. Hence the non-linear term, proportional to $\as^2$  cannot produce the logarithm $\log(k^2/Q_0^2)$, to be contrasted with the linear term, that yields the leading $\as \log(k^2/Q_0^2)$ contribution. Thus we get the same  conclusion as from the analysis in the Mellin moments space: the non-linear correction in the BK equation enters at lower order of the logaritmic $\log(Q^2)$ resummation of the perturbative series, than the corresponding term in the GLR equation. This difference may be traced back to the logaritmic integration needed to obtain the collinear gluon distribution from the unintegrated one. As a direct consequence, we expect that for $t \gg \abar y \gg 1$, the non-linear corrections from BK equation are significantly weaker than in the GLR equation.

{
It should be interesting to revisit the original argument about the connection between the GLR equation and the BK equation in the double logartithmic approximation (DLA) given in Ref.\ \cite{Kovchegov:1999yj}, that has lead to a different conclusion than ours. That analysis was performed in the transverse position representation. Two key approximations were applied: (i) the dipole kernel was approximated by the leading behavior for large daughter dipoles, $r \ll r', |\mathbf{r}'-\mathbf{r}|$,
\be
{\mathbf{r}^2 \over {\mathbf{r}'}^2 (\mathbf{r}'-\mathbf{r})^2} \; \longrightarrow \;
\theta(r'-r) {\mathbf{r}^2 \over {\mathbf{r}'}^4 },
\label{eq:kov-approx}
\ee
where the parent and daughter size dipole vectors are given by $\mathbf{r}$  and
$\mathbf{r}', \mathbf{r}' - \mathbf{r}$ respectively, $\theta$ is the Heaviside function. In addition, consistently, $N(x,\mathbf{r}' - \mathbf{r})$ was approximated by $N(x,\mathbf{r}' )$. Furthermore, (ii): the DLA relation between the dipole cross section and the collinear gluon distribution,
\be
{N(x,\mathbf{r}) \over r^2} = {\as \pi^2 \over 2\Nc S_{\perp}}  xg(x, 1/r^2),
\label{eq:xgdla}
\ee
 was employed. This lead to the following approximate form of the BK equation:
 \be
 x{\partial \over \partial x}xg(x,1/r^2) = {\abar \over 2} \int_{r^2}  ^{1/\Lambda^2}
 {d{r'}^2 \over {r'}^2} \left[ 2 xg(x,1/{r'}^2)  -  {\pi^2 \alpha_s \over 2 N_c S_{\perp}} {r'}^2 (xg(x,1/{r'}^2))^2\right],
 \label{eq:kov-glr}
 \ee
where $\Lambda$ is a nonperturbative energy scale of QCD or alternatively the saturation scale. From this equation the GLR equation was obtained by taking the logartithmic derivative, $r^2 \partial / \partial r^2$. Unfortunately, step (i) is not accurate enough to provide the correct dependence on $r^2$ of the non-linear term.  In Eq.\ (\ref{eq:kov-glr}) the $r^2$ dependence of the r.h.s.\ is coming entirely from the lower end-point region of the ${r'^2}$ integration. This is justified for the linear term in the DLA due to its $1/{r'}^2$ leading behavior in the integrand, but is not correct for the non-linear term. It is clear when one considers the non-linear term in the BK equation rewritten in terms of $xg(x, 1/r^2)$, using (\ref{eq:xgdla}). It reads:
\be
- {\as^2 \over 4S_{\perp} }
\int^{r' < 1/\Lambda} d^2\mathbf{r}'  \,
xg(x,1/{r'}^2) \, xg(x,1/(\mathbf{r}' - \mathbf{r})^2).
\ee
When this more accurate expression is differentiated with respect to $\log(r^2)$, the result:
 \be
- {\as^2 \over 4S_{\perp} }
\int^{r' < 1/\Lambda} d^2\mathbf{r}'  \,
xg(x,1/{r'}^2) \, r^2{ \partial \over \partial r^2} xg(x,1/(\mathbf{r}' - \mathbf{r})^2)
\ee
receives contributions from the whole integration region of $\mathbf{r}'$ with no enhancement for $r' \to r$. Hence, in this integral all scales of the collinear gluon distributions are probed down to $\Lambda^2$, with no enhancement of a particular region from the integration kernel. So, the contributions of gluon distributions at different scales $1/{r'}^2$ are weighted only by the integration volume effects, and they enhance the large $r'$ region, corresponding to small  $1/{r'}^2 \sim \Lambda^2 $ scales\footnote{In fact, the evolution of gluon distributions introduces some enhancement of the large scales $1/{r'}^2$ and $1/(\mathbf{r}' - \mathbf{r})^2$ scales (i.e.\ the small $r'^2$ and $(\mathbf{r}' - \mathbf{r})^2$), but these effects are enhanced only moderately by powers of logarithms and cannot fully compensate the strong power enhancement of large $r'$ values due to the integration volume effects.}. This is different than in the case of analogous differentiation of approximate integral (\ref{eq:kov-glr}), where the scales in both gluon distributions go to large $1/r^2$. Of course, such modification of the scales in the gluon distributions changes the evolution lenght in the scale and consequently, also the powers of logarithms of the hard scale included in the final equation of the GLR type with respect to the exact BK equation. In particular, when the scale approaches $\Lambda^2$, the effects of evolution are not present and there are no contributions of the large scale logarithms. This point, however rather subtle, is essential. The scheme that we implemented is free from inaccuracies introduced by approximation (\ref{eq:kov-approx}).}

{
The BK equation describes the evolution of color dipole scattering amplitude. From this amplitude we obtain the underlying unintegrated gluon distribution that belongs to a wider class of Transverse Momentum Distributions (TMDs), see e.g.\ \cite{Angeles-Martinez:2015sea,vanHameren:2016ftb}. More precisely, the distribution $f(x,k^2)$ that we use is called the dipole TMD (in the TMD notation: $xG^{(2)}(x,k^2)$). In general TMDs are defined by expectation values of gauge link contours, that correspond to parton configurations, and therefore depend on the scattering process. An important TMD is called the Weizs\"{a}cker--Williams distribution
($xG^{(1)}(x,k^2)$ in the TMD notation), relevant for many physical processes, e.g.\ for dijet production in DIS. The Weizs\"{a}cker--Williams and dipole gluon TMDs have the same behavior for $k^2 \gg \Qs^2$, but differ dramatically for $k^2 \sim \Qs^2$ or $k^2 < \Qs^2$. They also obey different evolution equations: while the dipole gluon TMD is governed by the BK equation, in the evolution of the  Weizs\"{a}cker--Williams gluon TMD the quadrupoles play an important role and the evolution is more complicated \cite{Dominguez:2011gc}. Hence, in general, one may expect that the non-linear evolution equation of the integrated versions of different TMDs are also different. The present analysis is focused only on the dipole gluon TMD and the conclusions may be not applicable to other gluon TMDs and their integrated counterparts. Let us add that in the McLerran--Venugopalan model \cite{McLerran:1993ni,McLerran:1993ka} the the Weizs\"{a}cker--Williams TMD is directly related to $\phi(x,k^2)$ (see e.g.\ \cite{Dominguez:2011wm}), and in the double logarithmic approximation the evolution equation of $\phi(x,k^2)$ takes the GLR form. The equivalence of $xG^{(1)}(x,k^2)$ and $\phi(x,k^2)$, however, does not hold when the QCD evolution effects are taken into account. }

\subsection{Effects of the high gluon density regime}

{ Now we turn to effects of the non-linearity when the gluon density is large, in particular to the  gluon saturation regime. We shall perform a heuristic analysis of the impact of saturation effects. We choose a simple model of $f(x,k^2)$, that provides clear analytic insight. The key feature of the model is presence of $x$-dependent saturation scale, that plays the role of lower cutoff on $k$, separating the linear evolution region from the region with strong suppression effects due to unitarity corrections, which impose fundamental constraints on the dipole scattering amplitude. In addition, the geometric scaling property of the dipole cross section is assumed to hold. The geometric scaling was initially discovered in the HERA data for the total $\gamma^* p$ cross section \cite{Stasto:2000er}, and it holds with a good accuracy for color dipole cross section obtained from the BK equation at larger rapidities \cite{Levin:1999mw,Levin:2000mv,Golec-Biernat:2001dqn}. The scaling is equivalent to universality of shape of the dipole cross section at different rapidities, and is also known as the ``travelling wave'' phenomenon \cite{Munier:2003vc,Munier:2003sj}. The physics picture and generic features of the obtained results do not depend on details of the shape of $f(x,k^2)$. The dipole scattering amplitude obtained from the simple model we assume has the geometric scaling built in, and it is consistent with the unitatity constraints. The way the amplitude approaches the unitarity limit is slightly different from the Levin--Tuchin law \cite{Levin:1999mw} that follows from the BK equation, but this difference does not affect the conclusions of the analysis. A similar reasoning lead to an proposal to impose unitarity effects on QCD evolution at small~$x$ as an absorptive boundary \cite{Mueller:2002zm,Triantafyllopoulos:2002nz}, which is not sensitive to a particular way to approach the unitarity limit. The geometric scaling implies the following form of the unintegrated gluon density}
\be
f(x,k^2) = S_\perp \Qs ^2(x)\, h(k^2/ \Qs ^2 (x)),
\ee
where $\Qs (x)$ is the $x$-dependent saturation scale, and the function $h$ is the universal profile of the BK solution. BK equation leads to an approximate power dependence of the saturation scale, $\Qs^2(x) \simeq Q_0^2 (x_{\mathrm{in}}/x)^{\lambda}$, with $\lambda \simeq 0.3$. Unitarity of the color dipole cross section scattering of a very dense target implies the asymptotic behavior
\be
h(\xi) \simeq \xi^2\quad \mbox{for}\;\; \xi \to 0,
\ee
that corresponds to $f(y,k^2) \sim k^4$ for $k^2 \ll \Qs^2(y)$ \cite{Golec-Biernat:2001dqn}.
For $\xi \gg 1$  the behavior of $h(\xi)$ is driven mostly by the linear evolution. For the purpose of this analysis it is sufficient to approximate the large $\xi$ behavior of $h(\xi)$ by $\xi ^{\gamma_c}$, where $\gamma_c$ is a positive number, much smaller than~1, related to the anomalous dimension of the gluon distribution function.  The simplest model of $h(\xi)$ that incorporates both the features is
\be
h(\xi) = A  [ \xi^2 \theta(1-\xi) + \xi^{\gamma_c} \theta(\xi -1)],
\ee
where $\theta$ is the Heaviside function and $A$ is a numerical constant.

We apply this model to estimate the effect of the non-linear term in the BK equation on the collinear gluon distribution for various hierarchy of scales.  The regime of $Q^2 \gg Q^2 _s(y)$ was studied above in the double logarithmic limit. Using the model of the BK solution for $f(y,k^2)$ we get
\[
xg(x,Q^2) \simeq \int^{Q^2} {dk^2 \over k^2} f(x,k^2) =
A (x_{\mathrm{in}}/x)^{\lambda} \left[
\int^{\Qs^2(x)} {k^2 dk^2 \over \Qs^4(x)} + \int_{\Qs^2(x)} ^{Q^2} {dk^2 \over k^2} (k^2/\Qs^2(x))^{\gamma_c}\right]
\]
\be
= A  (x_{\mathrm{in}}/x)^{\lambda}  \, \left[1/2 + {(Q^2 / \Qs^2(x))^{\gamma_c} - 1\over \gamma_c}\right].
\ee
Applying the expansion in $\gamma_c$ around zero up to the first order, and keeping only the leading logarithmic term we get
\be
xg(x,Q^2) \simeq  (x_{\mathrm{in}}/x)^{\lambda}  \log( Q^2 / \Qs^2(x)).
\ee
In the absence of non-linear correction the saturation scale in  $\log( Q^2 / \Qs^2(x))$ should be replaced by a much smaller scale $\mu_0 \ll \Qs(x)$, giving
$xg(x,Q^2)|_{\mathrm{linear}} \simeq  (x_{\mathrm{in}}/x)^{\lambda}  \log( Q^2 / \mu_0^2)$. Hence the relative correction due to non-linearity reads
\be
{xg(x,Q^2) - xg(x,Q^2)|_{\mathrm{linear} }\over xg(x,Q^2)|_{\mathrm{linear}}} \simeq
{ \log( Q^2 / \Qs^2(x)) -  \log( Q^2 / \mu_0^2) \over \log( Q^2 / \mu_0 ^2)}
= - { \log (\Qs^2(x) / \mu_0 ^2) \over \log( Q^2 / \mu_0 ^2)}\, .
\ee
This correction enters without a suppressing power factor of $1/Q^2$, hence at the leading twist, and due to logarithmic dependencies on the scales it is not small. This is consistent with our findings of strong non-linear correction in the proton structure functions at twist~2.
The presented estimate is rather crude, but it clearly shows how the non-linear corrections contribute to twist observables. It happens because the gluon recombination / unitarity leads to a strong suppression of unintegrated gluon distribution $f(x,k^2)$ in the region $k^2 < \Qs^2(x)$, and this imposes an effective lower cut-off on the logaritmic integration in $xg(x,Q^2) = \int_{\Qs^2(x)} dk^2 / k^2 f(x,k^2)$.

In these consideration we assumed that the saturation scale $\Qs \gg \mu_0$, where $\mu_0$ should be interpreted as an intrinsic hadronic scale of the proton, for instance it could be related to the inverse proton size. This is not in the perturbative domain, but this does not endanger the conclusions as $\mu_0$ enters only as a lower cut-off of logarithmic integrations. It should be clearly distinguished from  an initial scale of the DGLAP evolution $\mufo$, which is typically set to be greater than 1~GeV, and greater than the saturation scale in the proton. For the case  $\mufo \gg \Qs(x)$ the leading twist non-linear correction enter mostly as the input of the DGLAP evolution, with weak correction terms, as described in Sec.\ \ref{Sec:NL-DLA}.

\subsection{The intermediate region}

Above we discussed the non-linear effects for $Q^2 \gg \Qs ^2(x)$ and in the high gluon density regime. The intermediate region of $Q^2 \sim \Qs^2(x)$ is hardest to analyze analytically, as in this region $\Qs^2(x) / Q^2 \simeq 1$ and the linear and non-linear effects have similar size. Also, the relevant scale logarithms are of order~1, and the double logarithmic approximation is expected to have a very limited accuracy there. Hence, we believe that for a reliable numerical predictions in this region should be obtained within a complete non-linear evolution framework as given e.g.\ by the complete BK equation or the JIMWLK equation. Measurements of proton structure functions from HERA and their phenomenological analysis suggest, that the bulk of HERA data, say for $Q^2 > 2$~GeV$^2$ are driven by the linear evolution regime, as the saturation scale at HERA is below 1~GeV. {{The situation is expected to be different for deep inelastic scattering on heavy nuclei, that is going to be studied at the Electron--Ion Collider and in the possible future Large Electron--Hadron Collider (LHeC) \cite{LHeCStudyGroup:2012zhm,LHeC:2020van}. For a nucleus with mass number $A$ one expects the saturation scale $\Qs^2$ to be enhanced by a factor of $A^{1/3}$ with respect to the proton case. Moreover at the LHeC one expects to probe the range of $x$ that may extend down to $2 \cdot 10^{-6}$ at $Q^2 = 1$~GeV$^2$, see e.g.\ \cite{Rojo:2009ut}.  Hence one may reach $\Qs^2$ of few GeV$^2$ and there are good prospects to probe the non-linear effects well in the perturbative regime --- at the lowest $x$ the LHeC data on the proton and nuclear structure functions should have $Q^2 < \Qs^2$. In this region one expects that the dominance of twist~2 does not hold and the DGLAP framework may be not sufficient to describe the data. Perhaps an even more remarkable scenario may be realized ---  given the fact that the twist expansion in our approach leads to an asymptotic series with the expansion parameter $\sim \Qs^2 / Q^2$, one may expect that the twist expansion ceases to make sense for $Q^2 < \Qs^2$. Therefore it should be extremely interesting to probe this kinematic region experimentally and investigate the theoretical side within the framework of non-linear evolution equations. The problem of potential impact of non-linear effects on the proton and nucleus structure functions at LHeC was addressed in several studies \cite{Rojo:2009ut,Accardi:2011qh,Marquet:2017bga,Armesto:2022mxy}. In particular, in a recent analysis \cite{Armesto:2022mxy} careful matching of the DGLAP and BK description was perfomed for $(x,Q^2)$ range where both approaches are valid, and the differences between the predictions at small~$x$ and moderate $Q^2$ were studied. On a general level, the numerical results found in Ref.\ \cite{Armesto:2022mxy} have a similar pattern to the one following from our study, which, however, is more focused on identifying the structural features, and not yet on precision fits to the data.
}}

\section*{Acknowledgements}

This research was supported by the Polish National Science Centre (NCN) grant no.\ 2017/27/B/ST2/02755.

\bibliographystyle{JHEP}
\bibliography{bkht}

\end{document}